\documentclass[aps,prl,14pt,twocolumn,superscriptaddress,groupedaddress]{revtex4}  
\usepackage[utf8]{inputenc}
\usepackage{amsmath}
\usepackage{amsfonts}
\usepackage{amssymb}
\usepackage{graphicx}
\usepackage{bm}        

\usepackage{dcolumn}   

\begin{document}

\newcommand{\braket}[3]{\bra{#1}\;#2\;\ket{#3}}
\newcommand{\projop}[2]{ \ket{#1}\bra{#2}}
\newcommand{\ket}[1]{ |\;#1\;\rangle}
\newcommand{\bra}[1]{ \langle\;#1\;|}
\newcommand{\iprod}[2]{\bra{#1}\ket{#2}}
\newcommand{\logt}[1]{\log_2\left(#1\right)}
\def\cI{\mathcal{I}}
\def\cC{\tilde{C}}
\newcommand{\cx}[1]{\tilde{#1}}

\def\be{\begin{equation}}
\def\ee{\end{equation}}

\title{Distribution of quantum discord in Heisenberg Antiferromagnets}
\author{Aritra Kundu}
\email{aritrakundu@gmail.com}
\author{V. Subrahmanyam}
\email{vmani@iitk.ac.in}
 \affiliation{ Department of Physics, Indian Institute Of Technology,  Kanpur-208016, India}
 \date{\today}
\begin{abstract}
The quantum discord, which quantifies the amount of quantum correlations present between parts of a system, is investigated for antiferromagnetic spin systems. The discord for a pair
of spins in the many-spin ground state is related to the diagonal and off-diagonal spin-spin correlation functions and the local magnetization. For
isotropic and translationally invariant states,  the discord is shown to be  a function of  the diagonal correlation function only. Thus, near a thermal/quantum critical point, the
discord for a pair of spins shows long-range behavior, analogously of the correlation function. 
The discord exhibits a kink singularity as a function
of the anisotropy parameter for the the ground state of the Heisenberg model, for both nearest-neighbor spins as well as for well-separated spins. The preferred measurement basis for the minimum
conditional entropy, which determines the discord, changes discontinuously across the critical point. The conditional entropy distribution over all possible the measurement basis is
investigated. For the isotropic model, the distribution is just a $\delta$-function, whereas it has a twin-peak structure for anisotropic model. It is shown that the average value
and the mean-square fluctuation  of the conditional entropy  also show a signature of  the critical-point behavior.
 
\end{abstract}
\maketitle

\section {I. Introduction}

Entanglement quantifies quantum correlations between various parts of the system,  and it has been recognized as a resource for quantum 
computational tasks\cite{neilsen}.   Quantum correlations act as markers for both thermal and quantum phase transitions, being able to 
indicate a critical point (CP) through singularities. In recent years the study of quantum 
correlations and Entanglement in spin 1/2 spin systems has gained importance due to their potential application in building quantum computers.  

For many-qubit spin systems, it is interesting to study phase transitions from the perspective of how  quantum entanglement and information are shared between different pairs of spins. In particular there are many well-studied spin models, for which the entanglement and information distribution can be studied easily,  as these systems have
known diagonal and off-diagonal correlation functions exhibiting critical behavior near a phase transition point\cite{plenio}.
Various measures of entanglement have been used to study phase transitions in spin systems \cite{modi12, subrah}. The spin-spin correlations become long-ranged in the vicinity of a
quantum or classical critical point. However, this may not translate to a long-ranged behavior of the entanglement, as the entanglement measures use a combination of both the diagonal and the off-diagonal spin-spin correlation functions\cite{subrah1}.

A quantum-information-theoretic measure, viz. the quantum discord, has been explored by Oliver and Zurek\cite{zurek}, and Vedral\cite{vedral}, which incorporates how measurements
on subsystems affect the quantum correlations shared between two subsystems.
They argued that the quantum discord quantifies the minimum amount of quantum correlations for mixed states,  thus the discord characterizes the quantumness of the system.  
It is shown  that a nonzero discord can 
provide quantum advantage over classical systems even when the system has zero concurrence measure \cite{ad_power}
The quantum discord  has been explored  in the recent years\cite{Dillen, sarandy, indrani} in the context of quantum phase transitions.
In this paper we study the behavior of quantum correlations and quantum discord near a critical point in general, and then explicitly study pairwise quantum correlations in Heisenberg spin chain. 
We will show that the quantum discord has a long range behavior if the two-point diagonal correlation function is long ranged,  for isotropic and translationally-invariant systems. Thus, the quantum discord can track the critical behavior of systems, near either  a classical thermal critical point  or a quantum critical point, in an analogy with
the spin correlation functions.

The paper is organized as follows. We first derive an analytic formula, in the next section, for the quantum discord for a class of two-qubit mixed states. This is an improvement to the previous formulas \cite{luo, ali} that this method completely characterizes the optimal measurement basis. In the third section  we study the discord in spin systems with different symmetries. For isotropic states, we show that the discord between any pair of spins is completely determined by their diagonal correlation functions.
We show that the discord exhibits a kink at the Kosterliz-Thouless transition for the anisotropic Heisenberg model. We investigate the behavior of
the discord with the distance between spins.  In the fourth section we investigate the distribution of conditional entropies  as the measurement basis is varied for the subsystem. Instead of just picking the minimum value of the conditional entropy, as is done while computing  the quantum discord,  a distribution of conditional entropies is associated with a given two-qubit mixed state.  We study the average conditional entropy and the mean-square fluctuations of  this distribution for  the ground state of the anisotropic Heisenberg  model.

\section*{II. Quantum Correlations and Discord}\label{sec:Discord}

We will first consider two-qubit mixed states, and study the information-theoretic measure of discord between two spins. This measure will be used later, in the next section, to study the quantum correlations between
a given pair of qubits in a particular many-qubit state, for example the ground state of a spin chain with anisotropic Heisenberg interactions between neighboring spins. In this section, we focus
directly on a pair of  spins,  which can be viewed as a bipartite system.

Let $\rho_{AB}$ be a general two-qubit state comprising of A and B qubits.
Let $\rho_{A(B)}$ denote the reduced density matrix of the qubit $A(B)$, which is obtained through a partial trace over the states of $B(A)$,  $\rho_A=Tr_B ~\rho_{AB}$. 
The von Neumann entropy of a density matrix is defined as $S(\rho)=-Tr~ \rho \log_2\rho$.
The mutual information shared between the two qubits can be defined in terms of von Neumann entropies of the composite system and the subsystems,  given by 
\be  \label{eqn:TMI} J(\rho_{AB})=S(\rho_A)+S(\rho_B)-S(\rho_{AB}) .\ee
For a pure two-qubit state, this is just twice the entanglement, as $S(\rho_{AB})$ is zero and the other two von Neumann entropies are equal. Classically,  the mutual information can be equivalently written in terms of the conditional entropy, $C(A|B)$,  using the Bayes theorem that relates the joint and conditional probabilities of a composite system,  $C(A|B)=S(A, B)/S(B)$. The mutual information is now given as 
$ \label{eqn:CMI0}I(A, B)=S(\rho_A)-C(A|B)$.
In classical information theory $C(A|B)$ denotes the entropy of A, after  knowing that B takes a particular value.  Quantum mechanically the conditional entropy  requires measurement on qubit B, by choosing a particular measurement basis. Thus, the mutual information can depend on the measurement basis.

Let us set up a measurement basis $|\tilde k\rangle$ for the qubit B as,
\begin{eqnarray}\label{eqn:basis}
\ket{\cx{0}}=\cos{\frac{\theta}{2}}~\ket{0}+\rm e^{i\phi} ~\sin{\frac{\theta}{2}}~\ket{1}
\end{eqnarray}
and $\ket{\cx{1}}$ which is orthogonal to $\ket{\cx{0}}$, where $\ket{0}$ and $\ket{1}$ are the eigenstates of $s^z$ operator.
The conditional density matrix of $A$ with  $B=B_{\cx{k}}$, with an associated probability $p_{\cx{k}}$, is found by a partial trace over $B$, and using the corresponding projection operator,  
\be \label{eqn:condDM} \rho_{A|B_{\cx{k}}}=\frac{1}{p_{\tilde k}} \;Tr_B\; \projop{\cx{k}}{\cx{k}}\rho_{AB} .\ee
The mutual information $I_{\theta,\phi}(\rho_{AB})$  now depends on measurement basis,  as the conditional 
entropy depends on the measurement basis. We have,
\be  \label{eqn:CMI} I_{\theta, \phi}(\rho_{A B})=S(\rho_A)-C_{\theta, \phi}(\rho_{A|B}), \ee
where $C_{\theta, \phi}(\rho_{A|B})$ is the conditional entropy given by,
\be \label{eqn:condentropy} C_{\theta, \phi}({\rho_{A|B}})=p_{\cx{0}} S(\rho_{A|B_{\cx{0}}})+p_{\cx{1}} S(\rho_{A|B_{\cx{1}}}).\ee 

This  mutual information takes various values as the measurement basis is varied, and it may
not be equal to the mutual information $J(\rho_{AB})$,  given in Eq.1, which just depends on the
von Neumann entropy of the subsystem.  Now, the  
quantum discord is defined as the minimum difference between $J(\rho_{AB})$ and $I_{\theta,\phi}(\rho_{AB})$ taken over all possible measurement basis, given by
\begin{eqnarray}\label{eqn:discordgeneral}
\displaystyle D(A,B)=\min_{(\theta, \phi)}\;C_{\theta, \phi}({\rho_{A|B}})-S(\rho_{AB})+S(\rho_{B}).
\end{eqnarray}
Taking the minimum value of the difference between Eq.\ref{eqn:TMI} and Eq.\ref{eqn:CMI} corresponds to maximizing the classical mutual information over all the measurement bases. Thus, the discord defined above measures the minimum amount of quantum correlations (information) between the qubits in the mixed two-qubit state.
The best possible measurement basis which minimizes discord is characterized by two optimal angles $\theta$,$\phi$.
We will see below that, for pure states, this  reduces to the entanglement between the two qubits $A$ and $B$, i.e. entropy of the subsystem $S(\rho_A)$.  

In section IV we will study the full distribution of the conditional entropy values as the measurement basis is varied. Though, the quantum discord concerns only with the minimum value of
the conditional entropy, various other characteristics of the conditional entropy distribution are also
capable of quantifying the quantum correlations of the two-qubit state. In fact,  we will see that the first and the second moments of the conditional entropy distribution  can track the quantum phase transitions.

\subsection*{Minimization of conditional entropy}
We consider two-qubit states with a density matrix,  expressed in the $s^z$ diagonal basis as
\be \label{mat:densitymatrix}
\rho_{AB} = \bordermatrix{~ & 	\ket{00} & \ket{01} &\ket{10} & \ket{11}\cr
              \bra{00}  & u          & 0     & 0       &   y^* \cr
              \bra{01}  & 0          & w_1     & x^*       &   0 \cr
              \bra{10}  & 0          & x     & w_2       &   0 \cr
              \bra{11}  & y          & 0     & 0       &   v \cr
              }
.\ee
These states are known as the X states, as the nonzero elements of the density matrix are so located  to resemble that alphabet.
The conditional density matrix $\rho_{A|B_{\cx{k}}}$ is obtained from Eq.\ref{eqn:condDM},we have
\be
\rho_{A|B_{\cx{k}}}=\frac{1}{Tr \rho_{A|B_{\cx{k}}}}\bordermatrix{~\cr  & a^2u+b^2 w_1 & z^* \cr
					& z &	a^2 w_2 +b^2 v	\cr
					},
\ee
where $a=cos(\theta/2)$ and $b=sin(\theta/2)$,  and $z=\cos(\theta /2)\sin(\theta /2)[xe^{i\phi}+ye^{-i\phi}]$.
The eigenvalues of $\rho_{A|B_{\cx{k}}}$ are given by,
\be \label{eqn:eigenvalue} \lambda_{\cx{k}}^{\pm}=\left(p_{\cx{k}} \pm{\sqrt{b_{\cx{k}}^2+4|z|^2}}\right)/{2p_{\cx{k}}},\ee where
\begin{eqnarray}
p_0=(u+w_2)\cos^2(\theta /2)+(w_1+v)\sin^2(\theta /2),\nonumber\\
p_1=(u+w_2)\sin^2(\theta /2)+(w_1+v)\cos^2(\theta /2),\\
b_0=(u-w_2)\cos^2(\theta /2)+(w_1-v)\sin^2(\theta /2),\nonumber\\
b_1=(u-w_2)\sin^2(\theta /2)+(w_1-v)\cos^2(\theta /2).\nonumber
\end{eqnarray}
We can see from the above, replacing $\theta$ by  $\theta+\pi$,  $(p_0,b_0)$ and $(p_1,b_1)$ get interchanged. This implies that the conditional entropy is symmetric about $\theta=\pi/2$ and the minimum and maximum values occur at $\theta=0$ and $\theta= \pi/2$ respectively. 
When $\theta=0$,  $C_{\theta, \phi}(\rho_{A|B})$  becomes independent of $\phi$. We have,
\begin{eqnarray}
\label{ce0_0}{C_{0, 0}}(\rho_{A|B})=-u \logt{\frac{u}{u+w_2}}-w_2\logt{\frac{w_2}{u+w_2}} \nonumber \\ 
-v \logt{\frac{v}{v+w_1}}-w_1 \logt{\frac{w_1}{v+w_1}} .\nonumber\\
\end{eqnarray}

For $\theta=\pi/2$ the conditional entropy is optimized for $\phi=\phi^*$, where \be \tan(2\phi^*)=-{{\rm Im}(xy^*)}/{{\rm Re}(xy^*)}.\ee
For $x=0$ and/or $y=0$, the conditional  entropy is independent of $\phi$, as can be seen from Eq.\ref{eqn:eigenvalue}. The conditional entropy can be written in terms of Shannon Binary Entropy function $H(p)=-p \logt{p} -(1-p)\logt{1-p}$. We have, 
\be \label{ce0_45}
{C_{90, \phi^*}}=H\left (\frac{1+\sqrt{(u-v+w_1-w_2)^2+4(x+y)^2}}{2}\right).
\ee
The eigenvalues of the composite density matrix $\rho_{AB}$ are
${(u+v \pm \sqrt{(u-v)^2+4|y|^2}})/{2}$, $({w_1+w_2 \pm\sqrt{(w_1-w_2)^2+4|x|^2}})/{2}$ and $\rho_B$ is diagonal with eigenvalues $u+w_2$ and $v+w_1$. Calculating the corresponding entropies, and using 
Eq.6, we can write a compact form for the discord as,
\be \label{discord}
D({A,B})=Min\{C_{0, 0}, C_{90, \phi^*}\}-S(\rho_{A, B})+S(\rho_B).
\ee
Thus we have completely characterized the choice of basis state for calculation of discord of X states. 
We now consider a few simple cases below,.

\underline{Werner-separable state}: In this state the off-diagonal terms of $\rho_{AB}$ are zero. From Eq \ref{ce0_0}, and Eq.\ref{ce0_45},
we see that the preferred basis is $\theta=0$.  It is easy to see that the conditional entropy  $C_{0, 0}({A|B})$ is equal to $S(\rho_{A, B})-S(\rho_{B})$,  making the discord equal to zero, which is the least possible value. As expected the discord is identically zero for states with no off-diagonal terms. 

\underline{{Pure state}}:
The most general two qubit pure state is given by
$\ket{\psi_{AB}}=a\ket{00}+b\ket{01}+c\ket{10}+d\ket{11}$.  We can write this state in Schmidt basis as
 $ \ket{\psi_{AB}}=\sqrt{p}\ket{\cx{0}}_A\ket{\cx{0}}_B+\sqrt{1-p}\ket{\cx{1}}_A\ket{\cx{1}}_B $ 
 where 
 $ p ={(1+\sqrt{1-|2(bc-ad)|^2})}/{2}$. The kets $\ket{\cx{0}}, \ket{\cx{1}}$ are linear combinations of $\sigma_z$ eigenkets,  and are functions of the coefficients $a, b, c, d$. The density matrix can be written in the form given in Eq.\ref{mat:densitymatrix} with $u=p,v=1-p$, $w_1=w_2=0$, $x=0$ and $y=\sqrt{p(1-p)}$. Since it is a pure state $S(\rho_{AB})=0$ and $S(\rho_B)=H(p)$. From Eq.\ref{ce0_45}, we find that ${C}_{90, 0}(A|B)=0$ in this case.  This reduces the discord  to the entropy of qubit B, viz. the entanglement between A and B,
\be
D_{pure}({A,B})=S(\rho_B)=H(p)
\ee


\section*{III. Distribution of discord in a many-qubit state}
Let us consider a spin-1/2 system of N spins.  Any N-qubit pure state can be written as
 \be \ket{\psi}=\sum_{s_1..s_N} c(s_1..s_N) \ket{s_1..s_N} \ee where $c(s_1.  .s_N)$ is the probability amplitude for N-qubit direct product basis state $\ket{s_1s_2...s_N}$.  The site variable $s_i$ denotes the eigenvalue of the operator $s_i^z$,  and it can take  values $s_i=\uparrow $ or $\downarrow $.
The two-site reduced density matrix is  \be \rho_{i, j}=Tr' \rho =Tr' \projop{\psi}{\psi}. \ee
where prime denotes tracing over all spins other than the sites $(i, j)$.
We use the two-qubit basis states $|\uparrow\uparrow\rangle$, $|\uparrow\downarrow\rangle$, $|\downarrow\uparrow\rangle,  |\downarrow\downarrow\rangle$ for representing the density matrix.
This form is natural to spin representation and is equivalent to  the basis $\ket{0}$  and $\ket{1}$ used in the last section.
For a system where $[\rho, \sum_i (-1)^{s_i^z}]=0$,  the states with even $N_\uparrow$ do not mix with the
states with odd  $N_\uparrow$ , where $N_\uparrow$ is the number of spin with $s_i=\uparrow$ in the many-qubit basis state. The two-site density matrix will be of the form of the X states given in Eq.7.

The nonzero elements of the two-site density matrix can be written in terms of the diagonal and the off-diagonal correlation functions, and the local magnetizations  of the system\cite{subrah1}. We have $ u, v=1/4+\Gamma^D_{ij}\pm \langle (s^z_i+ s^z_j)\rangle $ and $ w_1, w_2=1/4-\Gamma^D_{ij}\pm\langle (s^z_i- s^z_j)\rangle $ where \be \Gamma^D_{ij}=\langle s_i^zs_j^z\rangle,\ee and\be x=\Gamma^O_{ij}=\langle s^+_is^-_j\rangle \\,  y=\langle s^+_is^+_j\rangle.\ee
If the total z component of the spins $s^z$ of the state is conserved, viz. $[\rho, s^z]=0$,  then states with  different number of $N_\uparrow$ do not mix with one another, implying that $y=0$.   If the state is time reversal symmetric then the magnetization terms  are zero,  and $x$ is real. 
The equations Eq.\ref{ce0_0} and Eq.\ref{ce0_45} are  simplified in this case.  Let us represent $\Gamma^O_{ij}=k\Gamma^D_{ij}$.  We have simplified expressions for the conditional entropies, given by

\be C_{90, \phi^*}(\rho_{s_i|s_{j}})=H(\frac{1}{2}+k\Gamma^D_{ij}),  ~C_{0, 0}(\rho_{s_i|s_j})=H(\frac{1}{2}+2\Gamma^D_{ij}).  \label{eqn:discord_spin}\ee 

Since the Shannon entropy function $H(p)$ is maximum at $p=1/2$, in the above we see that the value of k determines which of the two quantities is minimum. For  $k > 2$,  $\theta=90$ is preferred measurement basis, and  for $k<2$ the preferred basis is $\theta=0$. It follows $C_{90,\phi^*}\ge C_{0,0}$ when $k\ge2$ and $C_{0,0}\ge C_{90,\phi^*}$ when $k\le 2$. The positivity of eigenvalues of density matrix sets limits for $\Gamma_{ij}^D$, given by   \be \label{eqn:range} 1/4(1-k)\le \Gamma^D_{ij} \le  1/4(1+k) .\ee 

Let us consider many-qubit states with translational invariance, and  arrange the spins in a closed chain. Here, we can fix the focused pair
of spins to be $s_1$ and $s_{1+r}$.  All pairs of spins with separation $r$ will have the same correlation functions, and thus the same discord,  $D(s_1,s_{1+r})$.
The discord now can be written  in terms of the diagonal correlation functions and k, we have 
\begin{eqnarray} \label{eqn:discordanisotropic}
 D(s_1,s_{1+r})=1-S(\rho_{AB})+H(\frac{1}{2}+2\Gamma^D_{1,1+r})+\nonumber \\\theta(k-2) [H(\frac{1}{2}+k\Gamma^D_{1,1+r})-H(\frac{1}{2}+2\Gamma^D_{1,1+r})].
\end{eqnarray}
Now the eigenvalues of $\rho_{AB}$ are $\frac{1}{4}+\Gamma_{1,1+r}^D$ (two-fold degenerate),$\frac{1}{4}\pm(k\mp 1) \Gamma_{1,1+r}^D$,
and the eigenvalues of $\rho_B$ are 1/2 and 1/2. For $\Gamma_{1,1+r}^D<<1/4(k+1)$, the leading order is given by $D(s_1,s_{1+r}) \approx 2(k^2+4)(\Gamma^D_{1,1+r})^2/\log{2}$.
If the state is also rotationally symmetric, i.e. belongs to S=0 subspace, then $\Gamma^O_{1,1+r}=2\Gamma^D_{1,1+r}$, that is $k=2$, for all dimensions as all directions are equivalent.  In this case no minimization is required as $C_{90,0}(\rho_{s_i|s_{j}})=C_{0,0}(\rho_{s_i|s_{j}})$.
The discord depends on the diagonal correlation function only, for this case of isotropic states, we have

\begin{eqnarray} \label{eqn:discordisotropic} D(s_i,s_{j})=1+3(\frac{1}{4}+\Gamma_{1,1+r}^D)\log_2(\frac{1}{4}+\Gamma_{1,1+r}^D)+\nonumber \\
(\frac{1}{4}-3\Gamma_{1,1+r}^D)\log_2(\frac{1}{4}-3\Gamma_{1,1+r}^D)+H(\frac{1}{2}+2\Gamma_{1,1+r}^D).\nonumber \\
\end{eqnarray}
Since in the isotropic state there is no preferred orientation,  as expected the discord is independent of measurement basis. The limiting form of the discord,  for $\Gamma_{1,1+r}^D<<1/12$,  varies as the
square of the diagonal correlation function, we have 
 \be\label{eqn:asypisotropic} D(s_1,s_{1+r})\approx \frac{16} {\log{2}} (\Gamma_{1,1+r}^D)^2.\ee 
 
 Now, we can examine the behavior of the discord as a function of the separation between the pair
 of spins, as it just depends on the diagonal correlation function. Near
 a critical point, the correlation function is expected to be long ranged,  which implies a long-ranged
 discord.
In the vicinity of a thermal/quantum critical point, the diagonal correlation functions is expected to be of
the form,   $\Gamma_{i,i+r}^D(t)= e^{-r/\xi} /r$ for large r; where $\xi$ diverges as $\xi_0|1-t|^{-\nu}$ in general,  and as $e^{\pi/\sqrt{t-1}}$ for a Kosterlitz-Thouless-type transition. Here, $t=T/T_c$  is the ratio of the temperature and the critical temperature($T_c$) for a thermal phase transition and a corresponding dimensionless tuning parameter for a quantum phase transition.  For a pair of nearby
spins, the behavior of the correlation function does not follow the above form. The nearest-neighbor spin correlation function is directly proportional to interaction energy (as in Heisenberg model we will consider in the next section)\cite{takahashi}. The specific heat being a derivative of the internal energy, its critical behavior is related to the behavior of $\Gamma_{1,2}(t)$.  The leading-order behavior of the correlation function can be written,  using the specific heat exponent $\alpha$, as
 \be \label{eqn:NNcorrelation}\Gamma_{1,2}^D(t) \approx \Gamma_{1,2}^D(1)-\left(\Gamma_{1,2}^D(1)-\Gamma_{1,2}^D(0)\right)|1-t|^{1-\alpha}\ee
 
\begin{figure}
\centering
\includegraphics[width=\linewidth]{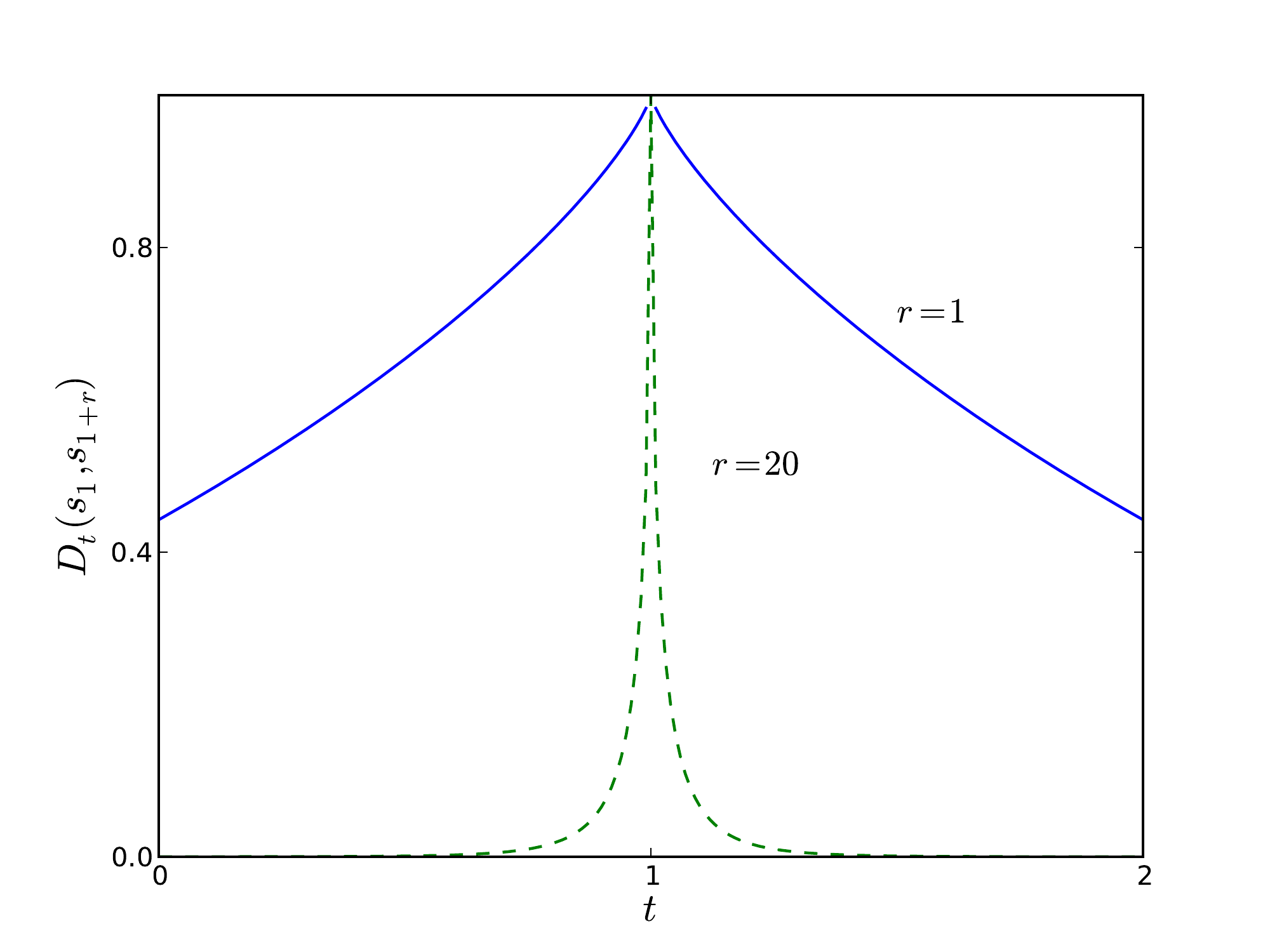}
\caption{The quantum discord $D_t(s_1,s_{1+r})/D_{t=1}(s_1,s_{1+r})$, for nearest neighbor spins and well-separated spins as a function of $t$ near a critical point ($t=T/T_c$ for a thermal critical point, and a corresponding tuning parameter for a quantum phase transition). The singularity in the nearest-neighbor discord, which uses the form for $\Gamma_{1,2}^D$ as given in Eq \ref{eqn:NNcorrelation}, is related to the specific heat exponent. The singularity for $r=20$ depends on the correlation length exponent.  } 
\label{fig:DiscordvsT_1}
\end{figure}

In the above equation, the first term  is just the value of nearest neighbor correlation function at the critical point, which we take as  $\Gamma_{1,2}(1)=-1/4$, for an antiferromagnetic system.  We estimate the
other constant  as $\Gamma_{1,2}^D(0)\approx -1/6$,  which varies about this value for other models. Therefore, $\Gamma_{1,2}^D$ is given by $-1/4(1-|1-t|^{1-\alpha}/3)$.  Using typical values of
the exponents,  $\alpha=0.1$, $\nu=0.6$ and $\xi_0=4$ (corresponding to a 3-dimensional Ising-type critical point\cite{huang}) , Fig.\ref{fig:DiscordvsT_1} shows the normalized discord, $D_t(s_1,s_{1+r})/D_{t=1}(s_1,s_{1+r})$ as a function of reduced temperature $t$, for correlations between nearest neighbor spins and  spins separated far apart($r=20$). Since the correlations for far-off spins are much smaller than nearest neighbors, the discord itself is very small for $r=20$ than nearest neighbors. In both cases, discord shows a similar behavior exhibiting a kink at the critical point. It is seen that as the system approaches criticality at $t=1$, the discord of the pair of spins attains its maximum value, which depends on the  separation between the spins.

\subsection*{Quantum Discord in anisotropic spin chains}
Consider a systems of spin-1/2 closed chain with $N$ spins with a nearest-neighbor anisotropic Heisenberg interaction,  with
the Hamiltonian given by
\be \label{eqn:hamiltonian} H=\sum_{i=1}^N(s_i^xs_{i+1}^x+s_i^ys_{i+1}^y+\Delta s_i^zs_{i+1}^z )\ee
where $\Delta$ is the anisotropy parameter. For $\Delta<0$, the coupling is ferromagnetic and for $\Delta>0$ it is antiferromagnetic in nature.  For $\Delta \leq -1$, it is energetically more favorable if the nearest neighbor spins are aligned in same direction.
For this Hamiltonian $s^z$ is good quantum number for all $\Delta$, i.e. it is time reversal invariant  and for $\Delta=1$, the total spin is also a good quantum number, thus the Hamiltonian is rotational invariant.  The ground state of the system belongs to $s^z=0$ subspace for $\Delta>-1$.
This mode is solved exactly using the Bethe Ansatz\cite{takahashi, orbach,walker}. 
The correlation function for $\Delta=1$ is expected to decay as $|\Gamma^D_{1,1+r}|\approx \
\sqrt{\log{r}}/r$. For $\Delta \ne 1$, the form of the correlation function is modified\cite{rpsingh}. 

\begin{figure} 
\includegraphics[width=0.5\textwidth]{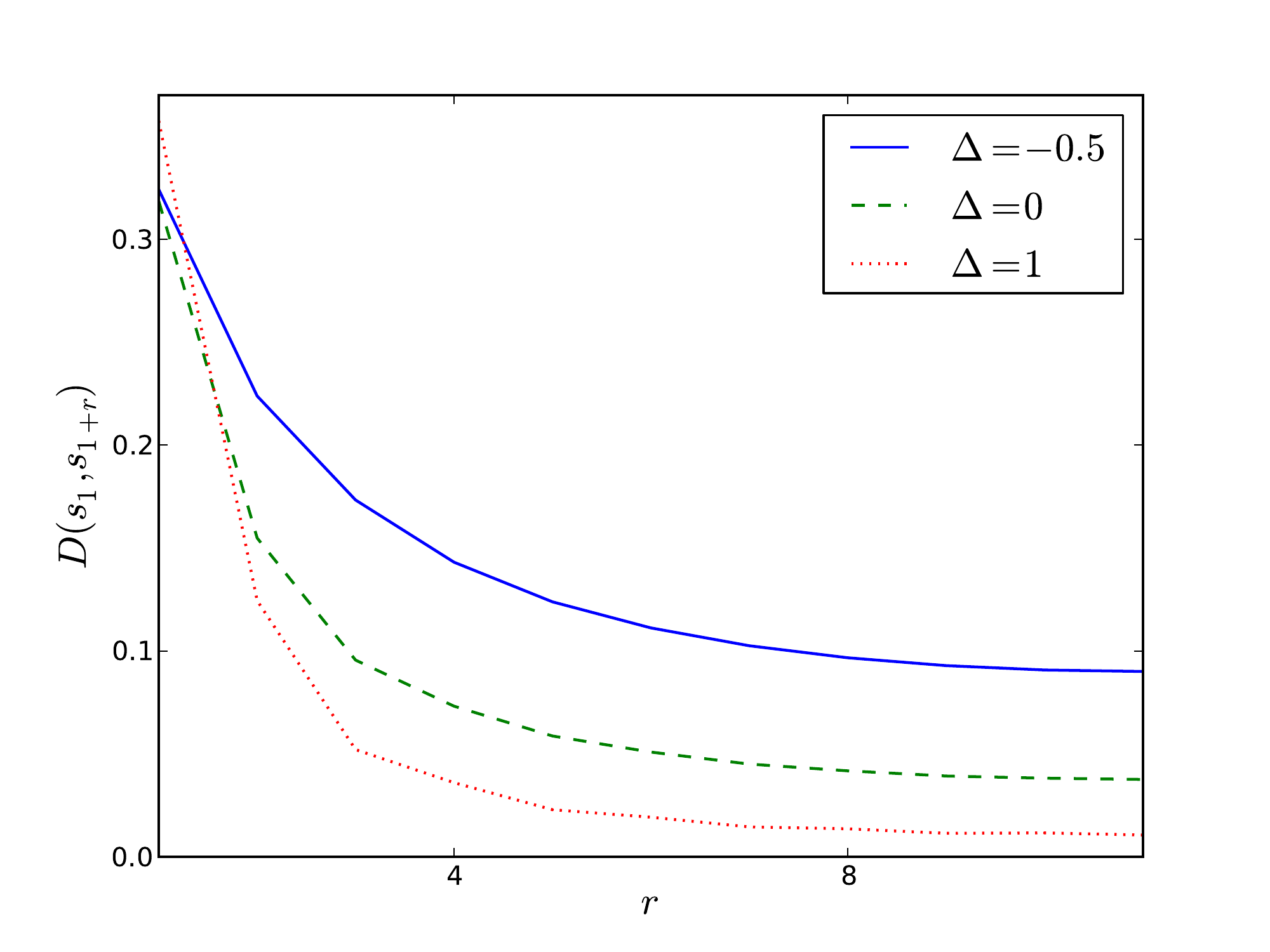}
\caption[Anisotropic discord model]{ The quantum discord $D(s_1,s_{1+r})$ is plotted as a function of the separation $r$ between the two spins, for a few values of the anisotropy parameter $\Delta$ calculated for $N=22$ spin chain with periodic boundary condition. For $\Delta<0$ (ferromagnetic case), the discord exhibits a slower decay. }
\label{fig:discordheisenberg14}
\end{figure}

Fig.\ref{fig:discordheisenberg14} shows the variation of $D(s_1,s_{1+r})$ as a function of the spin separation,  for different values of $\Delta$ for a closed spin chain with $N=22$ sites. 
The discord varies smoothly as a function of the distance between the spins in all values of $\Delta$. 
For $\Delta<0$, i.e. in the ferromagnetic regime,  $D(s_1,s_{1+r})$ is larger than that of the case when $\Delta>0$ for well-separated spins. This is because for $\Delta<0$, the large value of $k$ gives a
dominant contribution to the discord, which decays slower than the correlation function. At the critical point as the correlation length becomes infinite, the correlation function shows only a power-law decay, as $1/r$, the discord decays as $1/r^2$,.

\begin{figure}
\includegraphics[width=0.5\textwidth]{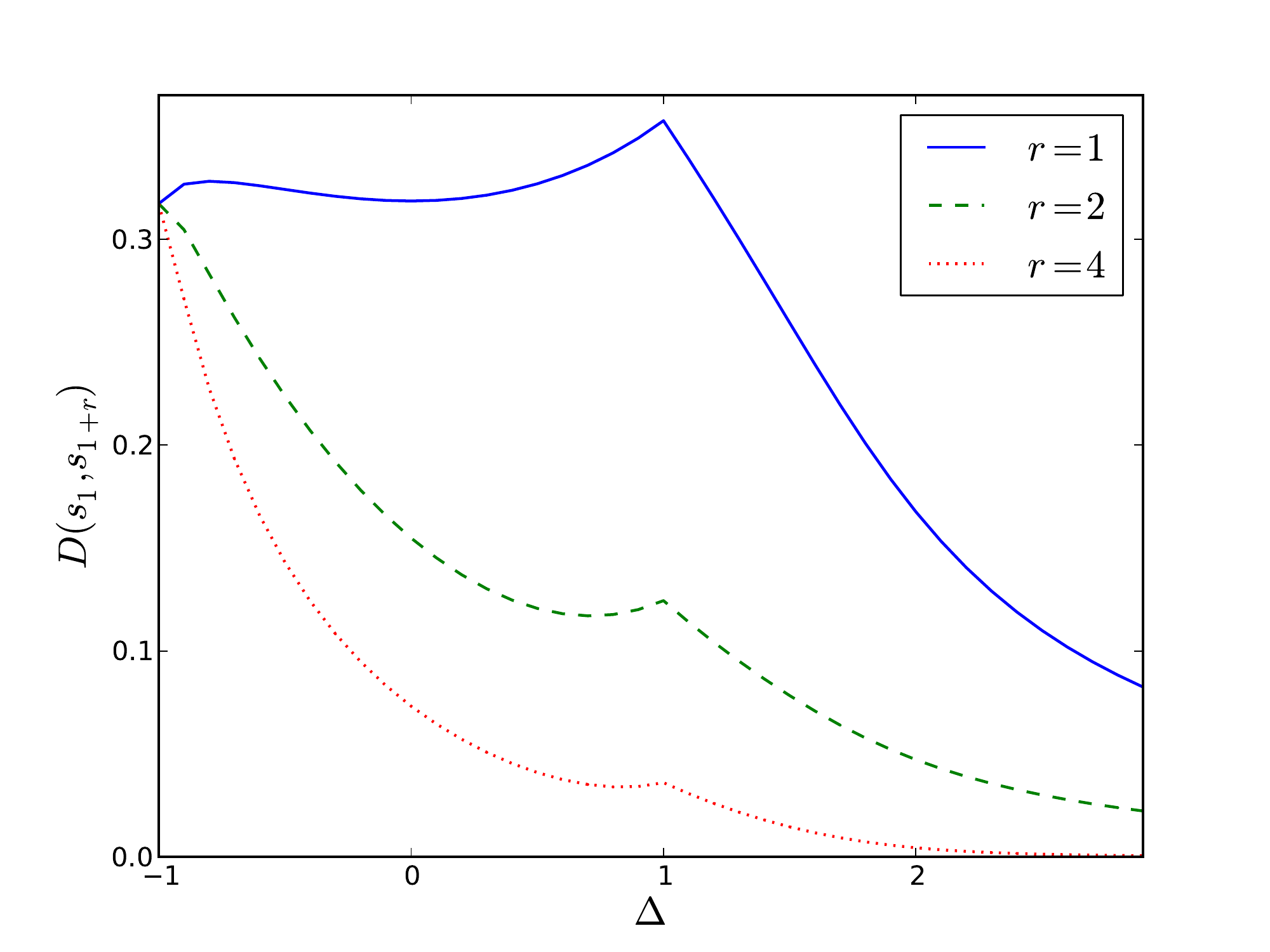}
\caption[Anisotropic discord model]{ The discord is plotted against the anisotropic parameter $\Delta$ from exact diagonalization for a $N=22$ closed spin chain, for separations $r=1,2,4,8$. For $\Delta<-1$ the discord is zero for all $r$. In all the cases, a kink singularity in discord at $\Delta=1$, similar to the one shown in Fig.\ref{fig:DiscordvsT_1}, tracks a quantum phase transition of Kosterlitz-Thouless type.}
\label{fig:anisotropydiscord}
\end{figure}

 
Fig[\ref{fig:anisotropydiscord}] shows the discord as a function of the anisotropy parameter $\Delta$, which exhibits a kink at $\Delta=1$ indicating a quantum critical point.  This transition corresponds to exponential correlation length decay as $\xi  =e^{\pi/\sqrt{\Delta-1}}$, valid close to the critical point from above,  which is characteristic of  a Kosterlitz-Thouless type transition\cite{kosterlitz,rpsingh}. 
Near the critical point, the variation of discord is dominantly decided by the factor $k$, as the correlation length diverges.
At the critical point,  the derivative of discord is discontinuous and the second derivative diverges. 
From Eq.\ref{eqn:discord_spin}, it can be seen that the discontinuity of discord at the quantum critical point is due to change in optimal basis from $s^x$ to $s^z$ from the transition in gapless($\Delta<1$) to gapped phase($\Delta>1$). We can argue that the maximum information about the orientation of the spins is obtained by measuring from a direction parallel to the polarization of the spins. 
The nearest-neighbor discord
$D(s_1,s_2)$ is largest for the isotropic case at $\Delta=1$, but for further-neighbor discord $D(s_1,s_{1+r})$ is maximum for $\Delta \approx -1$. At $\Delta\le -1$ since the state is ferro-magnetically ordered, $\Gamma^O=0$,  and hence the discord drops sharply to zero, indicating a phase transition. Thus, the discord can detect changes from gapless to gapped phases as it has the same length scale as the correlation length.

\begin{figure}
\includegraphics[width=\linewidth]{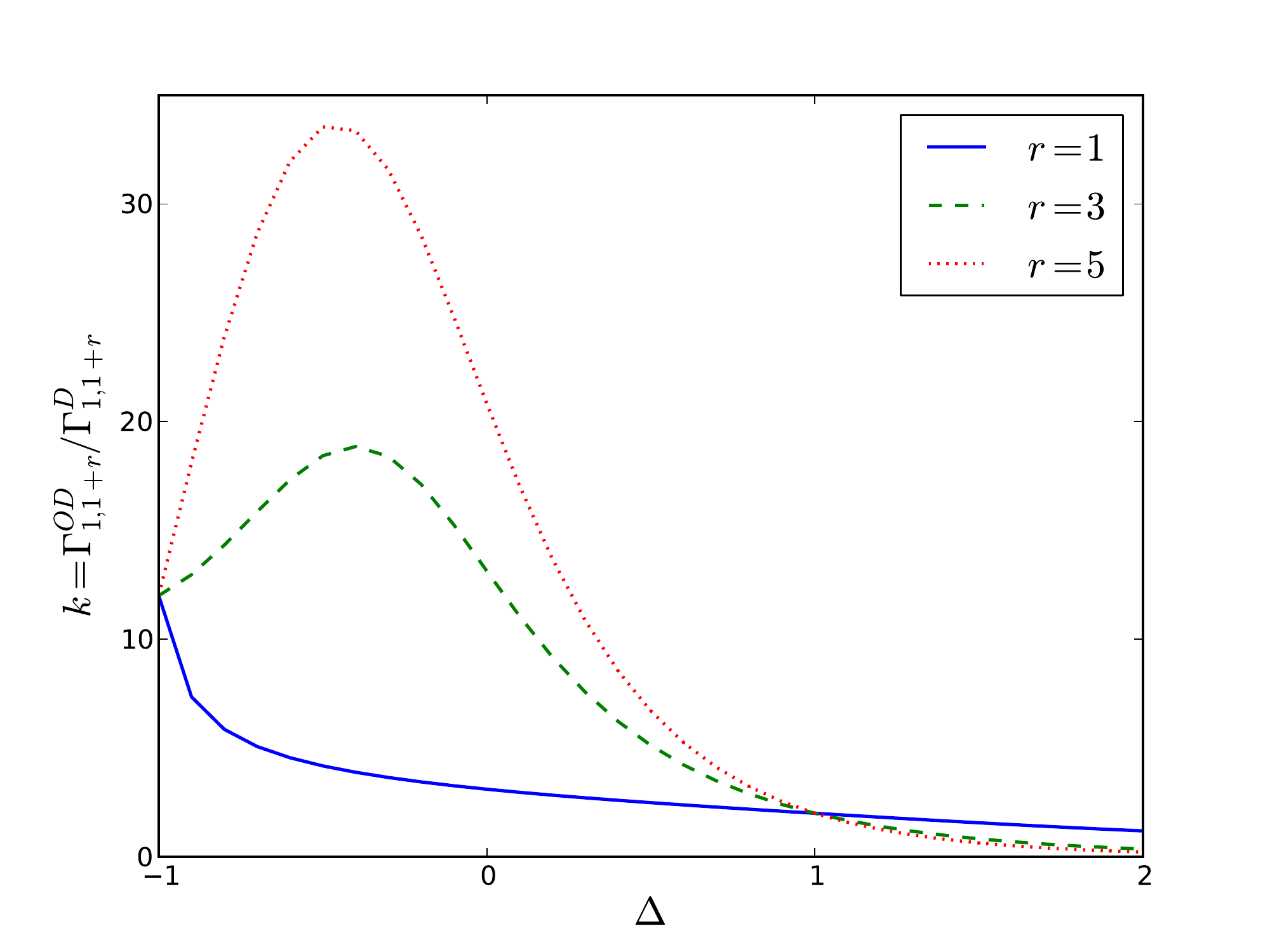}
\caption[variation of k with delta]{ The ratio of the off-diagonal and the diagonal correlation functions, $k=\Gamma^O_{1,1+r}/\Gamma^D_{1,1+r}$ for odd r is plotted as a function of the anisotropy parameter $\Delta$, for a $N=22$ closed spin chain. For all separations $r$, for $\Delta<1$, we have $k>2$. In this regime, the preferred basis for minimum conditional entropy Eq \ref{discord} is the $s^x$ basis. For $\Delta>1$, where $k<2$, the $s^z$ basis minimizes the quantum discord. A similar behavior is observed for even r. }
\label{fig:kvsdelta}
\end{figure}

In Fig.\ref{fig:kvsdelta} we plot the ratio of the correlation functions as a function of the anisotropy parameter $\Delta$. Since, $k=\Gamma^O/\Gamma^D$ determines the discord, the preferred measurement basis also depends on its value. From the numerical calculation for
$N=22$, it seems that for all spin separations, $k>2$ in the regime $\Delta <1$. 
In this case, the
preferred measurement basis is $s^x$ basis (or equivalently $s^y$ basis or any linear combination
of the two), as in this regime the interaction is XY-like. For $\Delta>1$, the Ising regime, we have
$k<2$, implying $s^z$ basis is preferred, In this case, spins are predominantly polarized along
the z axis. This behavior is captured by the discord as we discussed above.

\section{IV. Distribution of Classical Conditional entropy $C_{\theta,\phi}(\rho_{s_1|s_{1+r}})$}

The classical conditional entropy of a particular density matrix requires a selection of a measurement basis for one of the subsystems characterized by continuous parameters $\theta,\phi$.  
A particular value of the  conditional entropy may obtain for  multiple sets of parameter values.
We are hitherto concerned with the minimum value only, for computing the discord,  of the distribution of conditional entropies. To give a complete characterization of the state, a distribution of conditional entropies should be associated with a mixed two-qubit state, just as a mixed state should be viewed as a distribution over pure states. 
Apart from the minimum and the maximum values of the distribution, we can also study various moments of the distribution. The first and the second moments, which
are used to characterize a distribution in most practical situations, should also be able to track the
quantum correlations, as well as the discord that we studied in the previous section. As we will see below,  the average value and the mean-square deviation of the
conditional entropy distribution can also detect a critical point,
For each value of the conditional entropy $C$, we can associate a  probability $P(C)$, 
which is a measure of the likelihood of  the conditional entropy taking that particular value $C(\rho_{s_1|s_{1+r}})=C$. We define $P(C)$ as,
\be \label{eqn:P} P(C)= \frac{1}{4\pi} \int \delta (C-C_{ \theta,\phi}(\rho_{s_1,s_{1+r}}))d\Omega,\ee
where the integral is over all possible values of $\theta$ and $\phi$, thus spanning all the possible measurement bases. This distribution can be different for different pairs of spins in a given
many-spin state. We expect that the distribution will carry a strong dependence on the anisotropic parameter, as seen for the case of the discord. The full distribution of classical conditional entropy can capture more features of the conditional correlations than just the minimum of the distribution that has been used for calculating the discord.
For pure states, this distribution reduces to a Dirac-$\delta$ function at  $C=0$, and for Werner-separable states, the distribution becomes a Dirac-$\delta$ function at some particular value of $C$.

\begin{figure}
\includegraphics[width=\linewidth]{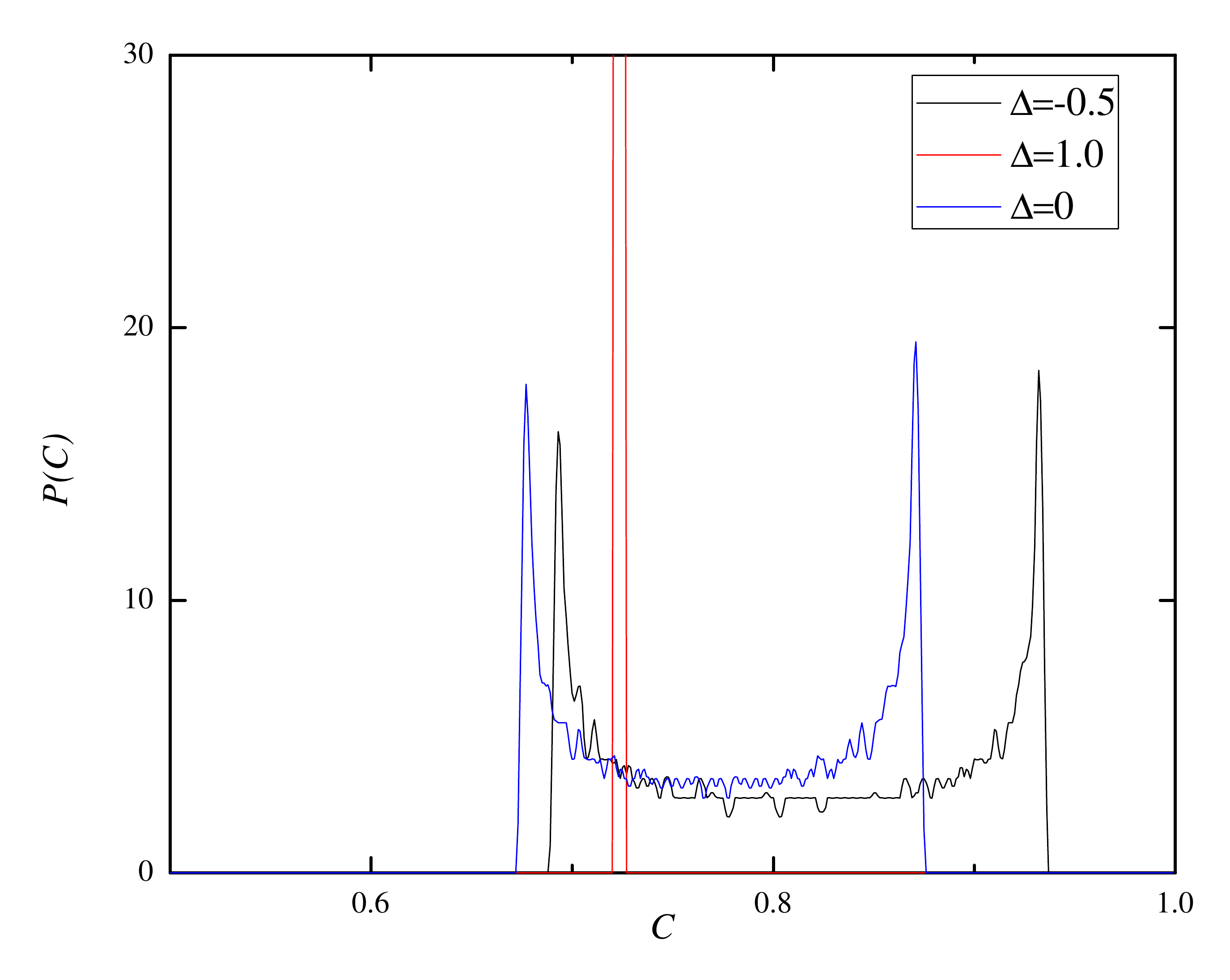}
\caption[Anisotropic]{ The probability distribution $P(C)$ is plotted against Conditional Entropy ($C$) for a 22-site closed chain for a few values of the anisotropic parameter $\Delta$. For $\Delta=1$, the distribution is just a delta function at around $C\approx0.72$. For $\Delta \neq 1$, the distribution has two peaks, the peak at lower value of $C$ is sightly less probable. However, the less probable peak determines the quantum discord. To show a smooth distribution, we have coarse grained the distribution with a bin size of $dC=0.005$.}
\label{fig:prob_dist_theta}
\end{figure}

We find that the distribution is difficult to obtain analytically in general,  though for the isotropic model is quite easy to construct. To this end, we
examine the conditional entropy distribution numerically, for a closed spin chain with $N=22$ spins, of a pair of spins with a given separation in the ground state of anisotropic Heisenberg model. Fig.\ref{fig:prob_dist_theta} shows the distribution of the nearest-neighbor conditional entropy for a  few values of the anisotropic parameter $\Delta$.  We have coarse-grained the distribution with a bin size of 0.005 to smoothen it. In the isotropic case,i.e. $\Delta=1$, since all the directions are equivalent, the distribution is independent of $\theta,\phi$, thus $P(C)$  just a $\delta$-function at $C=0.72$. For $\Delta\ne 1$, the distribution
shows a twin-peak structure, with peaks at the minimum and the maximum values of the conditional entropy,  with unequal probabilities.  The peak at the minimum value of $C$ is  more probable only
for  $\Delta\sim1$. 
The two peaks also correspond to the two orthogonal directions needed to specify the orientation of the spins,  corresponding to the two choices for $C_{\theta,\phi}$ that we discussed earlier (see Eq.20) .

\begin{figure}
\centering

\includegraphics[width=0.5\textwidth]{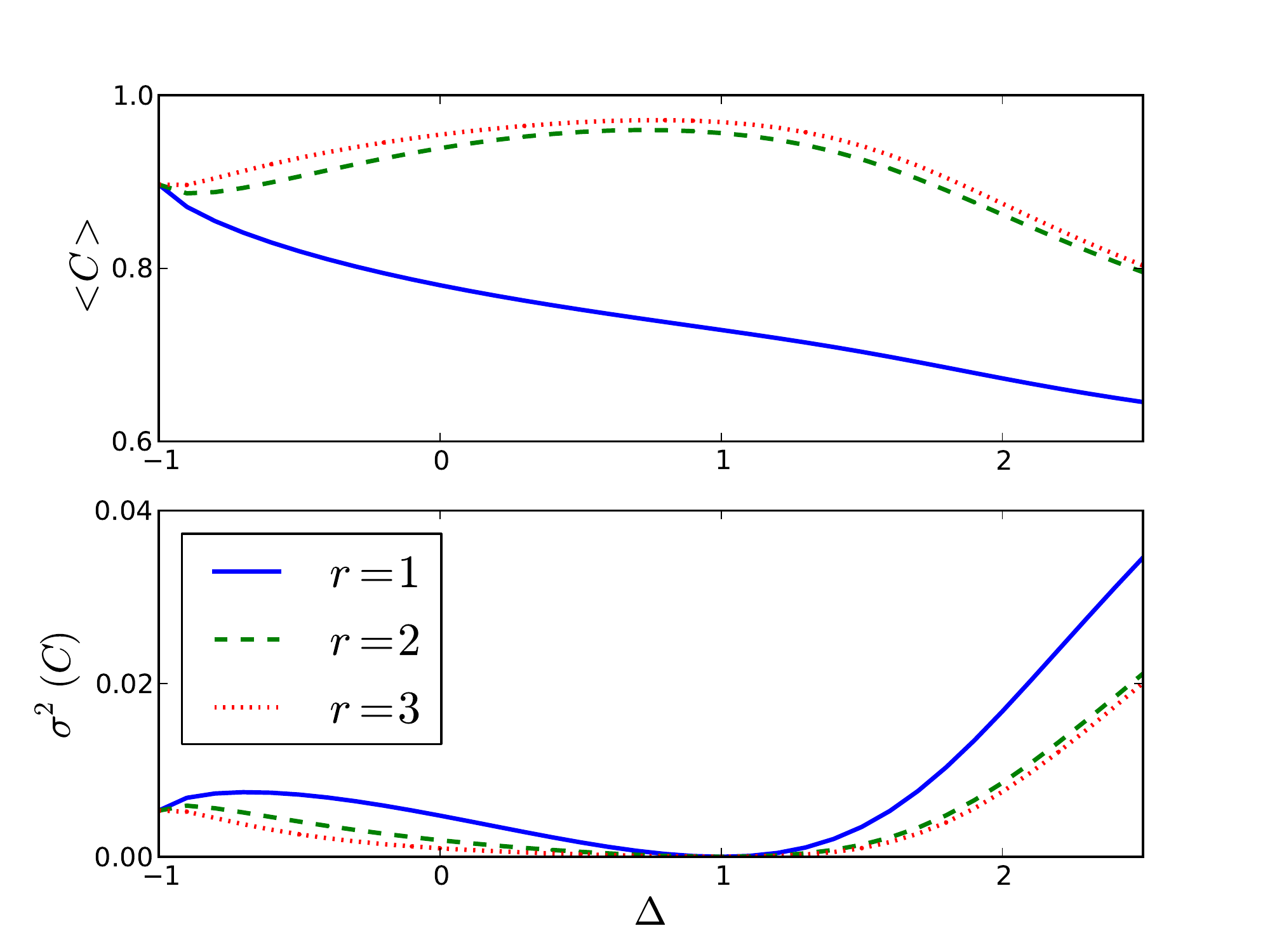}

\caption{The figure shows the average classical correlations $\langle C(\rho_{s_1|s_{1+r}})\rangle$ over all possible measurement basis for 22 sites closed chain anisotropic chain. The average value for pairs other than nearest neighbor shows a maximum near the critical point indicating criticality. Also note that the average value is as expected greater than the minimum value for all $\Delta$. The second figure shows the variation of square of standard deviation $\sigma^2(C)$ with delta. Clearly at quantum phase transition the distribution being delta function implies the standard deviation goes to zero at $\Delta=1$. }
\label{fig:Average_c_12_sites}
\end{figure}

Though, the quantum discord is defined using the minimum value of quantum correlations,  and analogously minimum of classical conditional entropy $ C_{\theta,\phi}(\rho_{s_1|s_{1+r}})$, 
in experiments it is more likely that we measure the average value of a quantity than either the minimum or most probable value. From the distribution we can calculate average and other moments of classical conditional entropy and hence the average quantum correlations. 
The average of the conditional entropy for Heisenberg model is shown in Fig.\ref{fig:Average_c_12_sites}. The average value of conditional entropy for nearest neighbors decreases with $\Delta$. For well separated spins, the average conditional entropy shows a maximum at $\Delta=1$, detecting phase transition. However, it is quite surprising that the nearest-neighbor average conditional entropy is not showing any structure at the critical point, but the further-neighbor
conditional entropy does show a peak structure.
We also show the mean-square deviation of the conditional entropy $\sigma^2(C)$  as a function of $\Delta$, in Fig.6, for different values of the separation. Here, for all separations the mean-square deviation shows a minimum at the critical point.  Clearly at a quantum phase transition the distribution being $ \delta$-function implies that the standard deviation goes to zero at $\Delta=1$, and away from the critical point the fluctuations grow, thus detecting the phase transition.

In conclusion, we have shown that the quantum discord, and the conditional entropy distribution and its moments  can track the critical behavior, as they depend on the spin correlations present
in many-spin ground state of anisotropic Heisenberg model.  A closed form of quantum discord is given in Eq.\ref{discord}, that is applicable for the two-qubit X states . The discord is directly determined by the diagonal correlation function, for rotationally and translationally symmetric states with
time-reversal invariance. Thus, the discord is long ranged, analogously as the correlation function exhibits long-range behavior in the vicinity of a thermal or quantum critical point. 
The quantum discord for
a pair of well-separated spins also is shown to track the Kosterlitz-Thouless type critical point in the anisotropic Heisenberg model. The conditional entropy for a pair of spins has a distribution of
values as the measurement basis for one qubit is varied. The conditional entropy distribution has been studied as a function of the anisotropic parameter. The average value and the mean-square fluctuation amplitude of the conditional entropy distribution exhibits a peak and valley structure respectively close to the critical point.


\end{document}